\documentclass[11pt,a4paper]{article}
\usepackage{jheppub}
\usepackage{subfig}

\def\half{\textstyle{1\over2}}

\newcommand{\del}{\ensuremath{\partial}}
\newcommand{\ba}{\begin{eqnarray}}
\newcommand{\ea}{\end{eqnarray}}

\newcommand{\be}{\begin{equation}}
\newcommand{\ee}{\end{equation}}

\title{Classical Duals, Legendre Transforms and the Vainshtein Mechanism}

\author{Antonio Padilla,}
\author{Paul M. Saffin}

\affiliation{School of Physics and Astronomy, University Park, University of Nottingham,\\ Nottingham NG7 2RD, UK}

\emailAdd{antonio.padilla@nottingham.ac.uk}
\emailAdd{paul.saffin@nottingham.ac.uk}

\abstract{We show how to generalize the classical duals found by Gabadadze {\it et al} to a very large class of self-interacting theories. This enables one to adopt  a perturbative description beyond the scale at which classical perturbation theory breaks down in the original theory. This is particularly relevant if we want to test modified gravity scenarios that exhibit Vainshtein screening on solar system scales.  We recognise the duals as being related to the Legendre transform of the original Lagrangian, and present a practical method for finding the dual in general; our methods can also be applied to self-interacting theories with a hierarchy of strong coupling scales, and with multiple fields. We find the classical dual of the full quintic galileon theory as an example.}


\begin{document}

\maketitle

\section{Introduction}
For any sensible theory of gravity, the dynamics  is described by a complicated coupled system of non-linear partial differential equations. This is true of General Relativity (GR), and must therefore be true of any theory that hopes to mimic GR at some appropriate scale. The complexity of the governing equations renders it difficult to find exact solutions, and typically one can only make progress by imposing certain symmetries in order to reduce the phase space, or else to consider perturbations about a known background solution.  Perturbation theory works best far away from the relevant excitation of the background, and will inevitably break down as we move closer and closer to the source. In  classical GR this breakdown occurs at the Schwarzschild radius of the  source. This is good enough for the most part, since we cannot experimentally probe the dynamics within the Schwarzschild radius since it would lie behind an event horizon.

In modified theories of gravity the situation can be more subtle. Large distance modifications of GR are often considered in order  to address an outstanding problem in cosmology such as the dark matter problem \cite{TeVeS}, the dark energy problem \cite{darkenergy}, and/or the cosmological constant problem \cite{CCP} (see \cite{review} for a general review and a more complete list of references).  This requires an ${\cal O}(1)$ deviation from GR on astrophysical or cosmological scales, and  for the theory to be phenomenologically viable, this deviation must reduce to $\lesssim {\cal O}(10^{-5})$ on solar system scales. Such dramatic suppression can sometimes be achieved through non-linearities, either from non-linear matter couplings, as in the chameleon mechanism \cite{cham}, or via non-linear  self-interactions, as in the Vainshtein mechanism \cite{vainshtein1, vainshtein2}; here we are primarily interested in the latter. 

Such self couplings have two important effects. At the quantum level, the interactions become strongly coupled above some particular scale, $\Lambda$, and one can no longer trust the classical background solution on distances $\lesssim \Lambda^{-1}$.  At the classical level, the interactions lead to the breakdown of the standard linearized theory around a heavy source.  In a generic modified gravity theory, with a non-relativistic source of mass $M$, the linearized perturbations  break down at the {\it Vainshtein scale},  $r_V$, which  typically takes the form \cite{dvalithm}
$$r_V \sim \left(\frac{M}{M_{pl}}\right)^\frac{1}{1+4(1-\alpha)} \Lambda^{-1}, \qquad 0 <\alpha<1$$
For the Sun, this scale must lie beyond the edges of the solar system, making it much larger than the Schwarzschild radius.  This is both a blessing and a curse.  If the Vainshtein mechanism is effective the non-linearities help to suppress the modifications of GR making the theory compatible with observation. However, in the absence of a  perturbative description below the Vainshtein scale, it is difficult to test any corrections to the leading order effect. 

Recently, Gabadadze {\it et al} \cite{duals} considered two examples of a classical theory with derivative self-interactions, each admitting a standard perturbative description {\it above} the relevant Vainshtein scale.  They then presented a ``classical dual"  of each theory, describing exactly the same physics, but admitting  a perturbative description {\it below} the Vainshtein scale. This is not a duality in the usual sense of strong versus weak coupling. Rather it is a classical analogue in the sense that the classical  expansion parameter, $r/r_V$, is inverted, with the two descriptions giving us the flexibility to do perturbation theory over a much larger range of scales,  provided the classical effective theory is valid.  This is reminiscent of Vainshtein's original approach to massive gravity \cite{vainshtein1} in which he made use of an expansion in $r_V/r$ above the scale $r_V \sim \left(\frac{r_s}{m^4}\right)^{\frac{1}{5}}$, and an expansion in $r/r_V$ below the scale $r_V$, where $m$ is the graviton mass and $r_s$ is the Schwarzschild radius of the source.  Indeed, this inversion of the expansion parameter was a big factor in motivating Gabadadze {\it et al}'s  recent work.

In this paper we give a general prescription for finding the classical duals of a much larger class of theories with self-interactions. Our methods work well for those theories that remain weakly coupled at low energies, and whose classical high energy dynamics is dominated by $N$-point interactions with finite $N$ only. We also require that the interactions become subdominant as the fields tend to zero.  

We begin by observing  that the classical duals presented in \cite{duals} actually make use of Legendre transforms of the interaction terms. Running with this idea, we are able to generalize their method. Indeed,  it is no coincidence that some mathematicians refer to the Legendre transform as  the {\it Legendre dual}\cite{ekland}. The duality is only useful if it admits a new perturbative description, and this is always true for the broad class of theories under consideration.
 
Whilst the Legendre transform picture is certainly instructive, it is not always straightforward to arrive  at a workable dual theory. This is down to technical difficulties in inverting the expression for the transformed variables. To alleviate this problem we will also present  a more practical method for finding the dual, using Lagrange multipliers, which are then integrated out. This should really  be considered as the working method for finding the classical dual of your favourite theory. It also helps us to identify exactly when a useful dual can be found. Both methods can be applied to theories with multiple scales, and multiple fields, in contrast to the examples given in \cite{duals}.

The rest of this paper is organised as follows: We start with a recap of the cubic galileon case covered in \cite{duals}, showing that the dual theory is just the Legendre transform of the original, and we then extend this idea in section \ref{sec:genLeg} to the general case, as well as proposing a practical implementation in section \ref{sec:pracMeth}. We see how models with multiple scales are dealt with in section \ref{multiscale} and multiple fields are examined in section \ref{multifield}. As an example of the method we work out the dual of the full galileon model in section \ref{sec:fullGal}, allowing for multiple scales, and then finish the paper with a discussion.

\section{The cubic galileon and its classical dual}
Let us begin by reviewing one of the examples considered by Gabadadze {\it et al} \cite{duals}, namely, the cubic galileon theory \cite{galileon}, 
\ba
{\cal L}&=&-\frac{1}{2}(\del \phi)^2-\frac{1}{\Lambda^3}\Box\phi(\del\phi)^2,
\ea
where  $\Lambda$ corresponds to the scale at which the interaction term becomes strongly coupled. This theory arises in the decoupling limit of the DGP model \cite{dgp, Luty, nic},  as well as in recent models of ghost-free massive gravity \cite{mg}.  Let us assume that matter couples to the scalar via an interaction of the form, 
\be
\phi \frac{T}{M_{pl}} \label{source},
\ee
where  $T$ is the trace of the energy momentum tensor. 
It follows that, classically, the linearised theory around a non-relativistic source of mass $M$ breaks down at the scale, $r_V \sim \left(\frac{M}{M_{pl}}\right)^{1/3} \frac{1}{\Lambda}$. To see this we note that the field equations can be schematically written as 
\ba
p^2 \phi -\frac{1}{\Lambda^3}(p^2 \phi)^2&\sim &\frac{T}{M_{pl}},
\ea
where $p$ corresponds to the momentum operator. For a non-relativistic source of mass, $M$, we get the right schematic behaviour by taking\footnote{Strictly speaking, for an approximately point-like source, we would take $T =-  M \delta^{(3)}(x)$ and integrate up the differential equations.  However, our schematic trick yields exactly the same behaviour, as of course it had to on dimensional grounds.} $T \sim M p^3$. In the linearised theory, $p^2 \phi \sim -\frac{T}{M_{pl}}$ and we obtain $\phi \sim \frac{M}{M_{pl}} p \sim  \frac{M}{M_{pl}} \frac{1}{r}$. It is clear that the linearised theory breaks down when $p^2 \phi \sim \Lambda^3$, and is only valid for  $r \gg r_V$. Above the Vainshtein scale we can explore corrections to the leading order behaviour by expansion in $r/r_V$.

Gabadadze {\it et al} \cite{duals} proposed the following classical dual to the cubic galileon theory
\ba
{\cal L}'&=&-\frac{1}{2}(\del \phi)^2-b^\mu \del_\mu \phi -\lambda \Box \phi +\Lambda^{3/2} \sqrt{\lambda b^2}.
\ea
A significant feature of  the dual formulation is that  the strong coupling scale, $\Lambda$, enters with positive powers. Thus the dual Lagrangian is well defined even in the limit $\Lambda \to 0$, in contrast to the original Lagrangian.  We also note that the two interaction terms are related by a  Legendre transformation, 
\begin{equation}
  -\frac{1}{\Lambda^3}\Box\phi(\del\phi)^2 \to -\Lambda^{3/2} \sqrt{\lambda b^2},
\end{equation}
with $\del_\mu \phi$ conjugate to  $-b^\mu$ and $\Box \phi$ conjugate to  $-\lambda$. With this observation, it now seems natural to include the canonical kinetic term in the transformation. To this end we define, 
\ba
b_{(1)}^\mu&=&\frac{\del {\cal L}}{\del \;\del_\mu\phi}=-\del^\mu \phi \left(1+\frac{2}{\Lambda^3} \Box \phi\right) \label{b1},\\
b_{(2)}&=&\frac{\del {\cal L}}{\del\;\Box\phi}=-\frac{1}{\Lambda^3} (\del \phi)^2 \label{b2}.
\ea
The Legendre transform of the Lagrangian is given by
\ba
f(b_{(1)}, b_{(2)}) &=&b_{(1)}^\mu \del_\mu\phi+b_{(2)}\Box \phi-{\cal L}=\pm \Lambda^{3/2} \sqrt{-b_{(2)} b_{(1)}^2}-\frac{\Lambda^3}{2} b_{(2)}.
\ea
The ambiguity in the sign reflects the ambiguity in inverting the relations (\ref{b1}) and (\ref{b2}). This stems from the fact that the cubic galileon theory admits two distinct branches. The classical dual is now given by
\be
{\cal L}'=b_{(1)}^\mu \del_\mu\phi+b_{(2)}\Box \phi-f(b_{(1)}, b_{(2)}),
\ee
and describes the same physics.  In the presence of the source (\ref{source}), we obtain the following field equations
\ba
\del_\mu b_{(1)}^\mu-\Box b_{(2)} &=& \frac{T}{M_{pl}} \label{Teq}, \\
\del _\mu \phi \pm \Lambda^{3/2}\frac{b_{(2)} b_{(1)}{}_\mu}{\sqrt{-b_{(2)}b_{(1)}^2}} &=& 0, \\
\Box \phi \pm \Lambda^{3/2}\frac{ b_{(1)}^2}{\sqrt{-b_{(2)}b_{(1)}^2}}+\frac{\Lambda^3}{2} &=&0.
\ea
These can be solved order by order in an expansion in $\Lambda^{3/2}$. For a non-relativistic source of mass, $M$,  schematically we have 
\ba
p b_{(1)}+p^2 b_{(2)} &=& \frac{M}{M_{pl}}p^3, \\
p \phi +\Lambda^{3/2} \sqrt{ |b_{(2)}| } \frac{b_{(1)}}{|b_{(1)}|} &=&0, \\
p^2 \phi +\Lambda^{3/2} \frac{b_{(1)}}{\sqrt{ |b_{(2)}| } }+\Lambda^3 &=&0.
\ea
It follows that 
\ba { b_{(1)} } & \sim&  \frac{M}{M_{pl}}p^2+ \Lambda^{3/2}\sqrt{\frac{M}{M_{pl}}}p^{1/2}+\ldots, \\
b_{(2)}  &\sim& \frac{M}{M_{pl}}p + \Lambda^{3/2}\sqrt{\frac{M}{M_{pl}}}p^{-1/2}+\ldots,  \\
\phi &\sim&   \Lambda^{3/2}\sqrt{\frac{M}{M_{pl}}}p^{-1/2}+\ldots,
\ea
corresponding to a perturbative expansion in $\left(\frac{\Lambda}{p}\right)^{3/2}\left(\frac{M}{M_{pl}}\right)^{-1/2} \sim \left(\frac{r}{r_V} \right)^{3/2}$.

\section{A general theory with self-interactions and its classical dual}
\label{sec:genLeg}
We will now consider general theories involving self-interactions,  extending the ideas initiated in \cite{duals}.   In the interests of clarity we present our analysis for a general theory involving a single field, $\phi$, of any type (we suppress tensor indices) with interactions all becoming strong in the UV, at the same scale $\Lambda$. Our generalization can be applied to theories with multiple scales, and multiple fields, and we will sketch how this should be done in sections \ref{multiscale} and \ref{multifield}.

We start with the Lagrangian density for the field $\phi$ propagating on  Minkowski spacetime
\ba \label{lag}
{\cal L} (\del^{(k)} \phi)\equiv {\cal L}(\phi,\del_\mu\phi,\del_\mu\del_\nu\phi, \ldots),
\ea
emphasizing once again that we are suppressing tensor indices on the field -- $\phi$ does not have to correspond to a scalar. We  will assume that  the field is ``canonically normalised" in some appropriate way, and that the propagator scales like $1/p^2$ in the UV. The various interactions are characterised by the number of  fields involved in the interaction, $N$,  and the number of derivatives,  $D$, and will schematically have the form
$$
\frac{ \del^D \phi^N}{\Lambda^{D+N-4}},
$$
where $D+N > 4$. This follows from the fact that  we only consider theories that are always weakly coupled at low energies,  so our interactions should disappear in the limit $\Lambda \to \infty$.  We also require that the interactions become subdominant as $\phi \to 0$ so we have $N>2$. If we further assume that the field couples to a source $J$ via an interaction
$$\phi J,$$
then the field equations are schematically given by
\be
p^2 \phi+\sum_i \frac{1}{\Lambda^{D_i+N_i-4}}p^{D_i} \phi^{N_i-1} \sim J.
\ee
Now we can probe the low energy physics by taking $\Lambda \to \infty$, in which case we have a good linearised theory with  $p^2 \phi \sim J$.  To probe the high energy physics, we must take the opposite limit $\Lambda \to 0$. Which term dominates the dynamics? Naively one might expect the dynamics to be dominated by the interaction containing the largest power of $1/\Lambda$. However, things are a little more subtle than that. It turns out that the dynamics is dominated by the term (or terms) with largest $t$, where
\ba
t=\frac{D+N-4}{N-1}
\ea
To see this, suppose that the $j$th interaction dominates the dynamics at high energies. It follows that as $\Lambda \to 0$, $\phi \propto \Lambda^{t_j}$, since $J$ is independent of $\Lambda$. At the level of the equations of motion, the $k$th interaction now scales as 
$$
p^{D_k} \left(\frac{\Lambda^{t_j}}{\Lambda^{t_k}}\right)^{N_k-1}.
$$ 
Since $t_j \ge t_k$ for all $k$, this will not diverge as $\Lambda \to 0$.

Now, if the largest value of $t$ occurs at finite values of $N$, then it is possible to identify the dominant UV behaviour and to expand around it. This controls whether or not we can find a classical dual that admits a useful perturbative description.  
Therefore, assuming that 
\be
\textrm{$t=t_{max}$ for finite values of $N$ only} \label{exist}
\ee  
we proceed to dualize the theory. 

\subsection{The Legendre dual}
To dualize our general theory, we simply compute the Legendre transform for the Lagrangian. To this end we define, 
\ba
a_{(0)}&=&\frac{\del {\cal L}}{\del\phi},\\
a_{(1)}^\mu&=&\frac{\del {\cal L}}{\del \;\del_\mu\phi},\\
a_{(2)}^{\mu\nu}&=&\frac{\del {\cal L}}{\del\;\del_\mu\del_\nu\phi},\\
&\vdots& \nonumber
\ea
The Legendre transform of the Lagrangian is given by
\be
f(a_{(k)})=a_{(0)}\phi+a_{(1)}^\mu\del_\mu\phi+a_{(2)}^{\mu\nu}\del_\mu\del_\nu\phi+\ldots-{\cal L}.
\ee
The precise form of this depends on the inversion of the relation $a_{(k)}=\frac{\del {\cal L}}{\del\; \del^k \phi}$. This may be multivalued, as it was for the cubic galileon. In any event, choosing some particular branch for the inverse, the dual theory is given by
\ba
\label{eq:L'}
{\cal L}' &=&a_{(0)}\phi+a_{(1)}^\mu\del_\mu\phi+a_{(2)}^{\mu\nu}\del_\mu\del_\nu\phi+...-f(a_{(k)}),
\ea
 with the following  field equations
\ba
a_{(0)}-\del_\mu a_{(1)}^\mu +\del_\mu\del_\nu a_{(2)}^{\mu\nu} +\ldots &=& -J, \\
\label{eq:phi-a}
\phi&=&\frac{\del f}{\del a_{(0)}},\\
\label{eq:dphi-b}
\del_\mu\phi&=&\frac{\del f}{\del a_{(1)}^\mu},\\
\label{eq:ddphi-c}
\del_\mu\del_\nu\phi&=&\frac{\del f}{\del a_{(2)}^{\mu\nu}}, \\
&\vdots& \nonumber
\ea
Now the original Lagrangian ${\cal L}\left(\del^{(k)}\phi\right)$  is such that all interactions remain weakly coupled at low energies, in the limit $\Lambda \to \infty$. It turns out the dual Lagrangian is well behaved in the opposite limit, $\Lambda \to 0$. To see why, we note that schematically
$$
a_{(k)} \sim \sum_i \frac{1}{\Lambda^{D_i+N_i-4}}p^{D_i-k} \phi^{N_i-1} \sim \sum_i p^{D_i-k} \left(\frac{\phi}{\Lambda^{t_i}}\right)^{N_i-1},
$$
This is inverted to find that the dominant $\Lambda$ scaling is $\phi\sim\Lambda^{t_{max}}$, from which it follows that $\del^{(k)}\phi \sim \Lambda^{t_{max}}  O^{(k)} (a)$ where the operators $O^{(k)}$ remain well behaved in the limit $\Lambda \to 0$. The Legendre transform is therefore given by
\be
f(a) \sim  \Lambda^{t_{max}}  \left[\sum_k  a_{(k)}   O^{(k)} (a)- \frac{{\cal L}( \Lambda^{t_{max}}  O^{(k)} (a))}{\Lambda^{t_{max}}}\right].
\ee
Now ${{\cal L}( \del^{(k)}\phi})\sim \phi \sum_i p^{D_i} \left(\frac{\phi}{\Lambda^{t_i}}\right)^{N_i-1}$ and so 
$$ \frac{{\cal L}( \Lambda^{t_{max}}  O^{(k)} (a))}{\Lambda^{t_{max}}} \sim \sum_i  \left(\frac{ \Lambda^{t_{max}}  }{\Lambda^{t_i}}\right)^{N_i-1}{\cal F}_i  (O^{(k)} (a)).$$
This is well behaved as $\Lambda \to 0$. It follows that the Legendre transform is well behaved as $\Lambda \to 0$, provided $t_{max} \ge 0$. This is indeed the case since we know that the quadratic term  has  $t=0$. We therefore confirm our assertion that the dual theory is well behaved in the limit $\Lambda \to 0$.

We see that the Legendre transform causes the expansion parameter to be inverted, the original theory working best at large $\Lambda$, with the dual working best at small $\Lambda$.   There is a characteristic  scale depending on $J$ and $\Lambda$ that acts as a pivot about which the duality is performed.  This is  precisely what you mean by the Vainshtein scale in certain modified gravity scenarios. On one side of the pivot we have the standard linearised theory with corrections that go like negative powers of $\Lambda$. On the other side we have the leading order short distance dynamics with corrections going like positive powers of $\Lambda$.  The dual theory gives us the means to study the latter using ordinary perturbative methods. 

The dual theory describes exactly the same physics as the original theory. This is manifestly true when one is considering excitations due to a source. When one is interested in freely propagating modes care must be taken to perform a non-covariant decomposition of the conjugate variables\footnote{We thank Gregory Gabadadze for pointing this out.}, as emphasized in  \cite{duals}. 
\subsection{A practical method for finding the dual}
\label{sec:pracMeth}
Now, in general, one cannot explicitly invert the relation $a_{(k)}=\frac{\del {\cal L}}{\del\; \del^{(k)} \phi}$.  For example, this is already true for the full galileon theory, including quartic and/or quintic interactions. \cite{galileon}.  This makes it difficult to find the dual theory using the Legendre transform method. Fortunately, however, we may use Lagrange multipliers to arrive at an equivalent dual theory with the same useful properties; we will now describe that method.

Introducing some Lagrange multipliers, $\zeta_{(k)}$,  and  auxiliary fields, $A_{(k)}$, we begin with a new Lagrangian, 
\ba
{\cal L}''&=&\zeta_{(0)}(\phi-A_{(0)})+\zeta_{(1)}^\mu(\del_\mu\phi-A_{(1)\mu})+\zeta_{(2)}^{\mu\nu}(\del_\mu\del_\nu\phi-A_{(2)\mu\nu})+\ldots +{\cal L}(A_{(k)}).
\ea
This is obviously entirely equivalent to our starting Lagrangian, $\cal L$, given by (\ref{lag}). However, the equations of motion for $A_{(k)}$ now correspond to constraints that we can use to integrate out the Lagrange multipliers.  In particular, we find
\ba
\zeta_{(0)}&=&\frac{\del {\cal L}}{\del A_{(0)}},\\
\zeta_{(1)}^\mu&=&\frac{\del {\cal L}}{\del A_{(1)\mu}},\\
\zeta_{(2)}^{\mu\nu}&=&\frac{\del {\cal L}}{\del A_{(2)\mu\nu}}, \\
&\vdots& \nonumber
\ea
Plugging this back into the action  we obtain, 
\ba\label{eq:lagrangeMult}
{\cal L}''&=&\frac{\del {\cal L}}{\del A_{(0)}}\phi+\frac{\del {\cal L}}{\del A_{(1)\mu}}\del_\mu\phi
                         +\frac{\del {\cal L}}{\del A_{(2)\mu\nu}}\del_\mu\del_\nu\phi+\ldots \\\nonumber
    &~&+{\cal L}(A_{(k)}) -A_{(0)}\frac{\del {\cal L}}{\del A_{(0)}}-A_{(1)\mu}\frac{\del {\cal L}}{\del A_{(1)\mu}}
      -A_{(2)\mu\nu}\frac{\del {\cal L}}{\del A_{(2)\mu\nu}} -\ldots
\ea
This is now of the same form as (\ref{eq:L'}), but by using the variables $A_{(k)}$  instead of $a_{(k)}$ we are able to get an explicit expression for the dual  Lagrangian. Note that 
\ba
a_{(k)}&=&\frac{\del {\cal L}}{\del A_{(k)}},
\ea
which is difficult to invert in general.

As it stands, there is no guarantee that the Lagrangian ${\cal L}''$ is well behaved as $\Lambda \to 0$. We can fix this  by rescaling the auxiliary fields   $A_{(k)} = \Lambda^{t_{max}} \hat A_{(k)}$.  To see why this helps consider the generic interaction 
$$
\frac{ \del^D \phi^N}{\Lambda^{D+N-4}} =\Lambda^4 \prod_i\left(\frac{\del^{(i)}\phi}{\Lambda^{i+1}}\right)^{n_i} \subset {\cal L}(\del^{(k)}\phi),
$$
where $D=\sum_i in_i$ and $N=\sum_i n_i$. It follows that
$$\frac{1}{\Lambda^{(N-1)t}}  \prod_i A_{(i)} ^{n_i} \subset {\cal L}(A_{(k)}), \qquad   \frac{n_j}{\Lambda^{(N-1)t} A_{(j)} }  \prod_i A_{(i)} ^{n_i}   \subset \frac{\del {\cal L}}{\del A_{(j)}},
$$
where we recall that $t=\frac{D+N-4}{N-1}$. Rescaling our variables, we obtain 
$$\Lambda^{t_{max}}\left(\frac{\Lambda^{t_{max}}}{\Lambda^{t}} \right)^{N-1}  \prod_i \hat A_{(i)} ^{n_i} \subset {\cal L}(A_{(k)}), \qquad   n_j\left(\frac{\Lambda^{ t_{max}}}{\Lambda^{t}} \right)^{N-1} \frac{1}{\hat A_{(j)} }  \prod_i \hat A_{(i)} ^{n_i}   \subset \frac{\del {\cal L}}{\del A_{(j)}}.
$$
Since $t_{max} \ge t$ and $t_{\max} \ge 0$, it is clear that as long as we replace $A_{(k)}$ with $\hat A_{(k)}$, then both ${\cal L}(A_{(k)})$ and $\frac{\del {\cal L}}{\del A_{(j)}}$ are well behaved as $\Lambda \to 0$.  As a result, ${\cal L}''$ is also well behaved in this limit. 

Let us see how this method works when applied to the cubic galileon. Note that the canonical kinetic term has  $t=0$, while the interaction has $t=3/2$, so we have $t_{max}=\frac{3}{2}$.   Before rescaling we have
\ba
{\cal L}''&=&\left(-\frac{2}{\Lambda^3}A_{(2)}A_{(1)}^\mu-A_{(1)}^\mu\right)\del_\mu\phi-\frac{1}{\Lambda^3}A^2_{(1)}\Box\phi
        +\frac{1}{2}A^2_{(1)}+\frac{2}{\Lambda^3}A_{(2)}A_{(1)}^2.
\ea
If we set  $A=\Lambda^{3/2}\hat A$ we obtain the dual Lagrangian
\ba
{\cal L}''&=&\left(-2\hat A_{(2)}\hat A_{(1)}^\mu-\Lambda^{3/2}\hat A_{(1)}^\mu\right)\del_\mu\phi-\hat A^2_{(1)}\Box\phi
        +\frac{1}{2}\Lambda^3\hat A^2_{(1)}+2\Lambda^{3/2}\hat A_{(2)}\hat A_{(1)}^2,
\ea
which is clearly well behaved as $\Lambda \to 0$, as desired.
\subsection{Multiple scales} \label{multiscale}
Our method for dualising can be easily adapted to deal with theories with more than one strong coupling scale.  To illustrate how, we consider a theory that depends on two scales $\Lambda \ll \bar \Lambda$, which we may schematically write as,
\be
{\cal L} \sim-\frac{1}{2} (\del \phi)^2 +\sum_i \frac{1}{\Lambda^{D_i+N_i-4}}\del^{D_i} \phi^{N_i} +\sum_j \frac{1}{\bar\Lambda^{\bar D_j+\bar N_j-4}}\del ^{\bar D_j} \phi^{\bar N_j}.
\ee
We see that we have explicitly separated  the interactions into two families -- those that become strong at $\Lambda$, and those that become strong at $\bar \Lambda$. Again, in order to guarantee that all interactions remain weakly coupled at low energies we assume $D+N>4$, and $\bar D+\bar N>4$.  We also assume $N, ~\bar N >2$ in order to ensure that the interactions become subdominant as $\phi \to 0$. 

For a source interaction of the form $\phi J$, the equations of motion are schematically given by
\be
p^2 \phi+\sum_i \frac{1}{\Lambda^{D_i+N_i-4}}p^{D_i} \phi^{N_i-1} ++\sum_j \frac{1}{\bar \Lambda^{\bar D_j+\bar N_j-4}}p^{\bar D_j} \phi^{\bar N_j-1} \sim J.
\ee
To probe the low energy physics we simply take $\Lambda, ~\bar \Lambda \to \infty$, and truncate to the linearised theory, $p^2 \phi \sim J$. We can perturb about the leading order solution using inverse powers of $\Lambda$ and $\bar \Lambda$. At higher energies we now have two distinct regimes. We first encounter an {\it intermediate} regime by  taking $\Lambda \to 0$, and $\bar \Lambda \to \infty$.  In contrast, the {\it high energy} regime is obtained by taking both  $\Lambda, ~\bar \Lambda \to 0$.  The importance of multiple scales and multiple classical regimes was emphasized in general galileon theories in \cite{clare}

Let us first consider the intermediate regime. We can obtain corrections to the leading order behaviour in terms of positive powers of $\Lambda$ and negative powers of $\bar \Lambda$. Introducing both $t=\frac{D+N-4}{N-1}$ and $\bar t=\frac{\bar D-\bar N-4}{\bar N-1}$, it is clear from our previous discussion that the intermediate scale dynamics is dominated by the term (or terms) with largest $t$ (with the $\bar t$ playing no role). To see this explicitly note that the leading order behaviour has $\phi \sim \Lambda^{t_{max}}$.  At the level of the field equations, the other interactions go like
$$
p^D \left(\frac{\Lambda^{t_{max}}}{\Lambda^t}\right)^{N-1}, \qquad p^{\bar D }\left(\frac{\Lambda^{t_{max}}}{\bar \Lambda^{\bar t}}\right)^{\bar N-1}.
$$
Now since $t, ~\bar t \ge 0$, and $t_{max} \ge t$, it is clear that  none of these terms diverge as $\Lambda \to 0, ~\bar \Lambda \to \infty$.

It is hopefully now obvious how to obtain a suitable dual in the intermediate regime.   The key point is that the interactions 
$$
\frac{1}{\bar\Lambda^{\bar D+\bar N-4}}\del ^{\bar D} \phi^{\bar N}
$$
are already in the correct form to admit an expansion in terms of inverse powers of $\bar \Lambda$. Thus we leave these interactions alone, and focus on dualizing the truncated Lagrangian
\be
{\cal L}_{truncated} \sim -\frac{1}{2} (\del \phi)^2 +\sum_i \frac{1}{\Lambda^{D_i+N_i-4}}\del^{D_i} \phi^{N_i}.
\ee
This can be obtained using either Legendre transforms, or our``practical" method, with $t_{max}$ playing the same critical role as outlined in previous sections.  We denote the resulting ``truncated' dual as ${\cal L}_{truncated}'$ and note that it is well behaved in the limit $\Lambda \to 0$. Combining it with the  other interactions,  we obtain the following dual for the full theory
\ba
{\cal L}'  &=& {\cal L}_{truncated}'+{\cal L}-{\cal L}_{truncated},\\
&\sim &  {\cal L}_{truncated}'+\sum_j \frac{1}{\bar\Lambda^{\bar D_j+\bar N_j-4}}\del ^{\bar D_j} \phi^{\bar N_j}.
\ea
This theory is well behaved as $\Lambda \to 0$ and $\bar \Lambda \to \infty$, and beyond this limit we can expand our classical solution in positive powers of $\Lambda$ and negative powers of $\bar \Lambda$. In a modified gravity model with two Vainshtein radii\cite{clare}, $r_V$ and $\bar r_V< r_V$, this expansion will ultimately be equivalent to an expansion in $r/r_V$, and $\bar r_V/r$. The expansion works well in the intermediate regime $\bar r_V<r< r_V$.

We now turn to the high energy regime. Corrections to the leading order behaviour are now expressed in terms of positive powers of both $\Lambda$ and $\bar \Lambda$. Which term (or terms) dominates the dynamics? To answer this we must introduce $\bar t_{max}$, the largest of the $\bar t$'s, in addition to $t_{max}$, the largest of the $t$'s. The term that dominates the high energy regime depends on the ratio of $\Lambda^{t_{max}}$ and $\bar \Lambda^{\bar t_{max}}$ in the  limit $\Lambda, ~\bar \Lambda \to 0$ .  If $\Lambda^{t_{max}} / \bar \Lambda^{\bar t_{max}} \nrightarrow \infty$ then the dominant terms stem from  $\Lambda$-type  interactions, $\frac{ \del^D \phi^N}{\Lambda^{D+N-4}} $, with $t=t_{max}$. In contrast, if $ \bar \Lambda^{\bar t_{max}}/\Lambda^{t_{max}}  \nrightarrow \infty$ then the dominant terms stem from  $\bar \Lambda$-type  interactions, $\frac{ \del^{\bar D} \phi^{\bar N}}{\bar \Lambda^{{\bar D}+{\bar N}-4}}$, with $\bar t=\bar t_{max}$.

To see why such terms dominate when they do,   we consider the  two  possibilities separately. If $\Lambda^{t_{max}}/\Lambda^{\bar t_{max}}$ is not divergent, then the $\Lambda$-type interactions with $t=t_{max}$ dominate and we have $\phi \sim \Lambda^{t_{max}}$.  To verify that this is correct, we need to show that none of the other interactions lead to divergences in the field equations in the desired limit.  Indeed, at the level of the field equations, the interactions go like
$$
p^D \left(\frac{\Lambda^{t_{max}}}{\Lambda^t}\right)^{N-1}, \qquad p^{\bar D }\left(\frac{\Lambda^{t_{max}}}{\bar \Lambda^{\bar t}}\right)^{\bar N-1}.
$$
The first of these will certainly not diverge as $\Lambda, ~\bar \Lambda \to 0$ since $t_{max}\ge t$. The second term is more subtle. To see how it behaves we rewrite it suggestively as
$$
p^{\bar D }\left(\frac{\Lambda^{t_{max}}}{\bar \Lambda^{\bar t_{max}}}\right)^{\bar N-1} 
\left(\frac{\bar \Lambda^{\bar t_{max}}}{ \bar \Lambda^{\bar t}}\right)^{\bar N-1}.
$$
Since $\Lambda^{t_{max}} / \bar \Lambda^{\bar t_{max}}\nrightarrow \infty$ and  $\bar t_{max}\ge \bar t$ , it is clear that this will not diverge as $\Lambda, ~\bar \Lambda \to 0$ . 

Now consider the alternative scenario, in which, $ \bar \Lambda^{\bar t_{max}}/\Lambda^{t_{max}}$ is not divergent. Using an entirely analogous argument one can easily prove that the $\bar \Lambda$-type interactions  with $\bar t=\bar t_{max}$ dominate.

It should now be clear how to take arrive at the  dual theory in the high energy regime. One can simply take the Legendre transform of the full Lagrangian. The dominant terms in the expansion of the transform then depend on the ratios of $\Lambda^{t_{max}}$ and $\bar \Lambda^{\bar t_{max}}$ in the limit. In any event, the resulting theory is well behaved as $\Lambda, ~\bar \Lambda \to 0$, and admits a perturbative expansion in positive powers of $\Lambda$ and $\bar \Lambda$. 

In applying our ``practical" method, we need to be sure to rescale the $A_{(k)}$ appropriately, depending on the dominant interaction. In particular, if  $\Lambda^{t_{max}} / \bar \Lambda^{\bar t_{max}}$ is not divergent, we introduce  $A_{(k)} = \Lambda^{t_{max}} \hat A_{(k)}$, whilst if  $\bar \Lambda^{\bar t_{max}}/\Lambda^{t_{max}} $ is not divergent, we introduce  $A_{(k)} = \bar \Lambda^{\bar t_{max}} \hat A_{(k)}$. In both cases, the resulting dual theory is guaranteed to be well behaved as $\Lambda, ~\bar \Lambda \to 0$, and to admit a perturbative expansion in positive powers of $\Lambda$ and $\bar \Lambda$.  In a modified gravity model with two Vainshtein radii, $r_V$ and $\bar r_V <r_V$, the dual theory would lend itself to an expansion in both $r/r_V$ and $r/\bar r_V$.

The generalization of the ideas presented in this section to theories with even more scales should now be obvious.

\subsection{Multiple fields} \label{multifield}
We shall now explain how our method should be generalized to deal with more than one field.  We will assume a single strong coupling scale for brevity, so that the theory may be schematically written as
\be
{\cal L}\sim \sum_\alpha -\half (\del \phi_\alpha)^2+\sum_i\frac{\prod_\alpha \del^{D_{i, \alpha}} (\phi_\alpha)^{N_{i, \alpha}}}{\Lambda^{D_i+N_i-4}},
\ee
where the index $\alpha$ labels the field and $D_i=\sum_\alpha D_{i, \alpha},~N_i=\sum_\alpha N_{i, \alpha}$. We have that $D_i+N_i>4$,  as well as $N_{i, \alpha}\geq 0$ and $N_i>2$. The first condition is required  for the theory  to remains weakly coupled at  low energies, while the  two  latter conditions are required in order to guarantee that the interactions become subdominant as $\phi_\alpha \to 0$. We assume interactions of the form $\phi_\alpha J_\alpha$, so that the equations of motion can be schematically written as
\be
p^2 \phi_\beta + \sum_i \frac{N_{i, \beta}}{\phi_\beta} \frac{\prod_\alpha  p^{D_{i, \alpha} } \phi^{N_{i, \alpha} }}{\Lambda^{D_i+N_i-4} } \sim J_\beta .
\ee
Now, generically, if a particular interaction (or interactions) dominates the dynamics as $\Lambda \to 0$, then we expect
$\phi_\alpha$ to scale the same way for each value of $\alpha$\footnote{This is easily seen by taking the ratio of the $\beta=\beta_1$ and $\beta=\beta_2$ equations of motion.  Assuming the term with $i=j$ is dominant, we have $\frac{N_{j, \beta_1} }{N_{j, \beta_2} } \frac{\phi_{\beta_2} }{\phi_{\beta_1}}\sim \frac{J_{\beta_1}}{J_{\beta_2}}$. The right hand side is independent of $\Lambda$, and so neglecting the special case where some of the $N_{j, \beta} $ vanish, we conclude that $\phi_{\beta_1} \sim \phi_{\beta_2}$}. We then claim that generically the term with largest $t_i=\frac{D_i+N_i-4}{N_i-1}$ dominates the dynamics.  To prove this we must show that none of the other interactions will  give a divergent contribution to the equations of motion as $\Lambda \to 0$.  To this end, we note that if our claim is true,  each field scales as $\phi_\alpha \sim \Lambda^{t_{max}}$.  At the level of the equations of motion, the interactions will now go like
\be
 \frac{1}{\Lambda^{t_{max}} } \frac{\prod_\alpha  p^{D_{i, \alpha} } \Lambda^{t_{max} N_{i, \alpha} }}{\Lambda^{D_i+N_i-4} } =p^D \left(\frac{\Lambda^{t_{max}}}{\Lambda^{t_i}}\right)^{N_i-1},
 \ee
where we have used the fact that $D_i=\sum_\alpha D_{i, \alpha},~N_i=\sum_\alpha N_{i, \alpha}$.  Since $t_{max} \ge t_i$ and $N_i>1$ it is clear that this does not diverge as $\Lambda \to 0$.

It is now clear that generically  we can take the dual of this theory in complete analogy with the single field case. Again, $t_{max}$ plays a critical role, particularly when applying the ``practical" method.

\section{The full galileon theory and its classical dual}
\label{sec:fullGal}
As an example of our method, we now consider the full galileon theory
\be
{\cal L}=\sum_{n=1}^{n=4}\frac{\alpha_n}{\Lambda_{(n)}^{3n-3}} \phi \delta^{[\nu_1}_{\mu_1}...\delta^{\nu_n]}_{\mu_n}{\del^{\mu_1}\del_{\nu_1}}\phi
                            \ldots{\del^{\mu_n}\del_{\nu_n}}\phi,
\ee
where  each $\alpha_n ={\cal O}(1)$, and in principle we have a hierarchy of as many as three different scales, $\Lambda_{(n)}$, $n=2, 3, 4$. The interaction terms have $t_n=3\frac{n-1}{n}$ which will be important in establishing how to rescale our conjugate variables.  

Our aim is to find the dual theory that is well behaved as $\Lambda_{(n)} \to 0$. Adopting the practical method, we first arrive at an equivalent Lagrangian,

\ba
{\cal L}'&=&\sum_n\frac{\alpha_n}{\Lambda_{(n)}^{3n-3}} {\phi} \delta^{[\nu_1}_{\mu_1}... \delta^{\nu_n]}_{\mu_n}{A^{\;\;\mu_1}_{(2)\nu_1}}
                 \ldots {A^{\;\;\mu_n}_{(2)\nu_n}}\\\nonumber
   &~&+\sum_n\frac{n\alpha_n }{\Lambda_{(n)}^{3n-3}} A_{(0)}\delta^{[\nu_1}_{\mu_1}...\delta^{\nu_n]}_{\mu_n}{(\del^{\mu_1}\del_{\nu_1}\phi)}A^{\;\;\mu_2}_{(2)\nu_2}\ldots 
                 A^{\;\;\mu_n}_{(2)\nu_n} \\\nonumber
   &~&-\sum_n\frac{n\alpha_n}{\Lambda_{(n)}^{3n-3} } {A_{(0)}}\delta^{[\nu_1}_{\mu_1}...\delta^{\nu_n]}_{\mu_n}{A^{\;\;\mu_1}_{(2)\nu_1}}
                 \ldots {A^{\;\;\mu_n}_{(2)\nu_n}}.
\ea

We need to rescale the variables according to the rules outlined in section \ref{multiscale}. How we do this depends on the ratios of the following in the limit where $\Lambda_{(n)} \to 0$, 
\be
\Lambda_{(2)}^{t_2}, \qquad \Lambda_{(3)}^{t_3}, \qquad \Lambda_{(4)}^{t_4},
\ee
where we recall that $t_n=3\frac{n-1}{n}$.  Let us assume, that we are interested in the case where $\Lambda_{(n_*)}^{t_{n_*}}/\Lambda_{(n)}^{t_n}$ is not divergent in this limit for some particular choice of $n_*$ and for any $n$. Then we perform the rescaling  $A=\Lambda_{(n_*)}^{t_{n_*}}\hat A$
\ba
{\cal L}'&=&\sum_n\alpha_n\left( \frac{\Lambda_{(n_*)}^{t_{n_*}}}{\Lambda_{(n)}^{t_n}}\right)^n {\phi} \delta^{[\nu_1}_{\mu_1}... \delta^{\nu_n]}_{\mu_n}{\hat A^{\;\;\mu_1}_{(2)\nu_1}}
                 \ldots {\hat A^{\;\;\mu_n}_{(2)\nu_n}}\\\nonumber
   &~&+\sum_n n\alpha_n\left( \frac{\Lambda_{(n_*)}^{t_{n_*}}}{\Lambda_{(n)}^{t_n}}\right)^n \hat  A_{(0)}\delta^{[\nu_1}_{\mu_1}...\delta^{\nu_n]}_{\mu_n}{(\del^{\mu_1}\del_{\nu_1}\phi)}\hat A^{\;\;\mu_2}_{(2)\nu_2}\ldots 
                 \hat A^{\;\;\mu_n}_{(2)\nu_n} \\\nonumber
   &~&-\sum_n n \alpha_n \Lambda_{(n_*)}^{t_{n_*}} \left( \frac{\Lambda_{(n_*)}^{t_{n_*}}}{\Lambda_{(n)}^{t_n}}\right)^n {\hat A_{(0)}}\delta^{[\nu_1}_{\mu_1}...\delta^{\nu_n]}_{\mu_n}{\hat A^{\;\;\mu_1}_{(2)\nu_1}}
                 \ldots {\hat A^{\;\;\mu_n}_{(2)\nu_n}}.
\ea
Because we are assuming that $\Lambda_{(n_*)}^{t_{n_*}}/\Lambda_{(n)}^{t_n}$  does not diverge as we take the $\Lambda_{(n)} \to 0$, it follows that this dual action is well behaved in the limit.  

Assuming that matter couples as in the cubic galileon case, we have the following equations of motion in the dual theory, 
\ba\nonumber
&~&\sum_{n=1}^{n=4}\alpha_n(\lambda_{(n)})^n\delta^{[\nu_1}_{\mu_1}... \delta^{\nu_n]}_{\mu_n}
                        \hat A^{\;\;\mu_1}_{(2)\nu_1}\ldots\hat A^{\;\;\mu_n}_{(2)\nu_n}\\
&&+\sum_{n=1}^{n=4}n\alpha_n(\lambda_{(n)})^n\delta^{[\nu_1}_{\mu_1}... \delta^{\nu_n]}_{\mu_n}
                        \del^{\mu_1}\del_{\nu_1}\left(\hat A_{(0)}\hat A^{\;\;\mu_2}_{(2)\nu_2}\ldots\hat A^{\;\;\mu_n}_{(2)\nu_n}\right)
+J=0,\\
&~&\sum_{n=1}^{n=4}\alpha_n(\lambda_{(n)})^n\delta^{[\nu_1}_{\mu_1}... \delta^{\nu_n]}_{\mu_n}
      \left(\del^{\mu_1}\del_{\nu_1}\phi-\Lambda^{t_{n_\star}}_{(n_\star)}\hat A^{\;\;\mu_1}_{(2)\nu_1}\right)
             \hat A^{\;\;\mu_2}_{(2)\nu_2}\ldots\hat A^{\;\;\mu_n}_{(2)\nu_n}=0,\\\nonumber
&~&\sum_{n=1}^{n=4}n\alpha_n(\lambda_{(n)})^n\phi\delta^{[\rho\nu_2}_{\;\sigma\mu_2}... \delta^{\nu_{n}]}_{\mu_{n}}
             \hat A^{\;\;\mu_2}_{(2)\nu_2}\ldots\hat A^{\;\;\mu_{n}}_{(2)\nu_{n}}\\\nonumber
&&+\sum_{n=1}^{n=4}n(n-1)\alpha_n(\lambda_{(n)})^n\hat A_{(0)}\delta^{[\rho\nu_2}_{\;\sigma\mu_2}... \delta^{\nu_{n}]}_{\mu_{n}}
             \del^{\mu_2}\del_{\nu_2}\phi \hat A^{\;\;\mu_3}_{(2)\nu_3}\ldots\hat A^{\;\;\mu_{n}}_{(2)\nu_{n}}\\
&&-\sum_{n=1}^{n=4}n^2\alpha_n\Lambda_{(n_\star)}^{t_{n_\star}}(\lambda_n)^n\hat A_{(0)}\delta^{[\rho\nu_2}_{\;\sigma\mu_2}... \delta^{\nu_{n}]}_{\mu_{n}}
             \hat A^{\;\;\mu_2}_{(2)\nu_2}\ldots\hat A^{\;\;\mu_{n}}_{(2)\nu_{n}}=0,
\ea
where $\lambda_{(n)}=\Lambda^{t_{n_\star}}_{(n_\star)}/\Lambda^{t_{n}}_{(n)}$. We now find the background solution in the following high energy limit: $\Lambda_{(n)} \to 0$ with   $\lambda_{(n)}\to \begin{cases}1 & n=n_* \\  0 & n \neq n_* \end{cases}$. Taking the source to be $J\sim\frac{M}{M_{pl}}p^3$, as before, the background becomes
\ba
A_{(2)}\sim\left(\frac{M}{M_{pl}}\right)^{1/n_\star}p^{\frac{3}{n_\star}},\\
A_{(0)}\sim A_{(2)}p^{-2}\sim\left(\frac{M}{M_{pl}}\right)^{1/n_\star}p^{\frac{3-2n_\star}{n_\star}},\\
\phi\sim\Lambda_{(n_\star)}^{t_{n_\star}}A_{(2)}p^{-2}\sim\Lambda_{(n_\star)}^{t_{n_\star}}\left(\frac{M}{M_{pl}}\right)^{1/n_\star}p^{\frac{3-2n_\star}{n_\star}},
\ea
with the higher order terms being by expansions in $\Lambda^{t_{n_\star}}_{(n_\star)}$ and in  $\lambda_{(n)}^{n}$ for $ n \neq n_*$. These correspond to expansions in 
$$
\left(\frac{\Lambda_{(n_\star)}}{p}\right)^{t_{n_\star}} \left(\frac{M}{M_{pl}} \right)^{-t_{n_*}/3} \sim \left(\frac{r}{r_{n_*}}\right)^{t_{n_\star}}
$$ 
and
$$
\lambda_{(n)}^{n} p^{\frac{3(n-n_*)}{n_*}} \left(\frac{M}{M_{pl}}\right)^{\frac{n-n_*}{n_*}} \sim\left( \frac{r}{r_n}\right)^{\frac{3(n_*-n)}{n_*}}, \qquad  n \neq n_*
$$
respectively. The theory admits up to three critical radii given by
$$r_n \sim \begin{cases}  \left(\frac{M}{M_{pl}}\right)^{1/3}\Lambda_{(n_\star)}^{-1}  & n=n_* \\
\left(\frac{M}{M_{pl}}\right)^{1/3} \lambda_{(n)}^\frac{nn_*}{3(n-n_*)} &n \neq n_*
\end{cases}
$$
thereby generalising the Vainshtein radius for the multiscale theory, as expected \cite{clare}. 
\section{Discussion}
Whenever classical perturbation theory breaks down at some particular scale, to  continue making predictions beyond that scale one would like to have a dual theory. This should describe the same physics, but admit a perturbative description that works best in the opposite regime to that in the original theory. This is exactly what Gabadadze {\it et al} \cite{duals} achieved by identifying the duals to two particular classical theories that exhibit Vainhstein screening. In each case, perturbation theory works best above the Vainshtein radius in the original theory, and below the Vainshtein radius in the dual.

In this paper, we have recognised these examples as being nothing more than the Legendre transform of the original Lagrangian. This has enabled us to generalize the idea, and outline how one can find the classical dual for a much larger class of self-interacting theories. We have also presented a more ``user-friendly" method for finding the dual for the case where it is difficult to compute the Legendre transform explicitly.

Of course, a dual is only of any use if it admits a complementary perturbative description, as in the examples given in \cite{duals}. If the original theory admits a good perturbative description as some dimensionful scale $\Lambda \to \infty$, this amounts to the dual theory admitting a good perturbative description in the opposite limit $\Lambda \to 0$. That this should happen places certain restrictions on the form of the theory one can successfully dualize. The key ingredient is  that there must exist a finite  interaction (or interactions) that dominates the classical dynamics as $\Lambda \to 0$, and about which we turn the theory on its head. This logic applies even when there are multiple scales or multiple fields.

We have presented the classical dual of the full galileon theory as an example.  Of course, there are many more examples one could consider, and one can find their duals using the methods we have discussed. We have also shown explicitly that a useful dual can only be found if certain  specific criteria are met: namely that there exists a finite $N$-point interaction (or interactions) with largest $t$, where $t=\frac{D+N-4}{N-1}$ and $D$ is the number of derivatives. 

Let us end with  an example of a Lagrangian for which we {\it cannot} find a useful classical dual.  Consider the Einstein-Hilbert action, written in terms of the metric expanded around Minkowski space, $g_{\mu\nu}=\eta_{\mu\nu}+\frac{1}{M_{pl}} h_{\mu\nu}$. Schematically we have
\be
S_{EH}=\int d^4 x ~h \del^2 h +\sum_{i=3}^\infty \frac{\del^2 h^i}{M_{pl}^i}.
\ee
As is well known, this theory is well behaved as $M_{pl} \to \infty$, as the  interactions vanish. Each interaction has $D_i=2$ derivatives, $N_i=i$ fields and so $t_i=\frac{D_i-N_i-4}{N_i-1}=-\frac{i+2}{i-1}$. Recall that the dynamics as $M_{pl} \to 0$ is dominated by the term (or terms) with largest $t_i$. However, this does not occur at a finite value of $i$, and so we cannot find a classical dual that is well behaved as $M_{pl}\to 0$.  This does not mean that one cannot find a classical dual to GR. Such a dual can be found but only if we describe the dynamics using something other than $h_{\mu\nu}$. This is currently a work in progress \cite{grdual}.

\section*{Acknowledgements}
We thank Gregory Gabadadze for a number of insightful comments.
AP is supported by a Royal Society University Research Fellowship, and PMS by STFC. 

\end{document}